\documentclass{article}
\usepackage{amssymb}
\newcommand{\be}{\begin{equation}}
\newcommand{\ee}{\end{equation}}
\newcommand{\bea}{\begin{eqnarray}}
\newcommand{\eea}{\end{eqnarray}}
\newcommand{\eqref}[1]{(\ref{#1})}
\begin{document}
\begin{center}
   \setlength{\baselineskip}{15pt} 
\textbf{{\large  STRING REPRESENTATION OF THE DUAL 
GINZBURG-LANDAU THEORY BEYOND THE LONDON LIMIT}}%
\footnote{Presented by M.~Koma at ``Confinement V'',
Gargnano, Italy, 10-14 Sep. 2002}
\\[+.9cm]
M.~KOMA\( ^{\mathrm{A,B}} \),
Y.~KOMA\( ^{\mathrm{A}} \), D.~EBERT\( ^{\mathrm{C}} \)
AND H.~TOKI\( ^{\mathrm{B}} \)
\\
\textit{\( ^{\mathrm{A}} \)Max-Planck-Institut
f\"ur Physik, M\"unchen\\
\( ^{\mathrm{B}} \)Research Center for Nuclear Physics(RCNP), 
Osaka University, \\
\( ^{\mathrm{C}} \)Institut f\"ur Physik, Humboldt Universit\"at zu Berlin \\
E-mail: mkoma@mppmu.mpg.de}\\[+.7cm]
\end{center}
\begin{abstract}
The effective string action of the color-electric flux tube  
in the 
dual Ginzburg-Landau (DGL) theory is studied 
by performing  a path-integral analysis by taking into account
the finite thickness of the flux tube.
A modified Yukawa interaction appears as a  boundary 
contribution and is reduced into the ordinary Yukawa
interaction in the London Limit.\\[+.4cm]
\end{abstract}

\setlength{\baselineskip}{13pt}
The dual Ginzburg-Landau (DGL) theory 
can sketch the dual superconductor scenario of quark confinement
mechanism intuitively
by the formation of a flux tube due to the dual Meissner effect.
Since the flux tube corresponds to
the hadronic object, the construction of
the effective string action of the flux tube
is one of the interesting applications of the DGL theory
to the hadron physics. 
In this brief report, 
we discuss the structure of the effective
string action of the $U(1)$ version of the DGL
theory by using path-integral analysis.\cite{Koma:2001pz}
In particular, we pay attention to the effect of the
finite thickness of the flux tube to 
the form of the string action.

The $U(1)$ DGL action in the differential form 
has the following form:%
\begin{eqnarray}
S_{\rm DGL} 
=
\frac{\beta_{g}}{2}  (F)^{2}
+((d-iB)\chi^{*}, (d+iB)\chi)
+\lambda (|\chi|^{2}-v^{2})^{2},
\label{eqn:weyl-dgl}
\end{eqnarray}
where  $B$ is the  dual gauge field (1-form)
and 
$\chi \! =\!\phi \exp (i\eta)$ ($\phi$, $\eta$ $\in \Re$)
the  complex-scalar monopole field (0-form).
The dual gauge coupling, the strength of the self-coupling and the
monopole condensate are denoted by $\beta_{g} \!=\! 1/g^{2}$, $\lambda$
and $v$, respectively. They are related to the two mass scales of the
theory:
the mass of the dual gauge field  $m_{B} \!= \!\sqrt{2}\, gv$ and 
that of the monopole field  $m_{\chi} \!=\! 2
\sqrt{\lambda}\,v$.
The inverses of these masses correspond to the penetration depth and
coherence length, respectively.

In the presence of external quark sources,
the dual field strength $F$ (2-from) is expressed as 
\(
F\! = \!dB\! -\! 2 \pi * \Sigma.
\)
In a dual model, an external $q$-$\bar{q}$ source is described as an
{\it open} color-electric Dirac string $\Sigma$ (2-form),
whose boundary gives a quark current $j$ (1-form).
By the Hodge decomposition,
the dual gauge field is decomposed into two parts as
$B \!=\! B^{\mathrm{reg}} \!+\!
B^{\mathrm{sing}}, $
where $
B^{\mathrm{sing} } \!= \!2\pi \Delta^{-1}\! \delta * \Sigma $.
Here, $\Delta^{-1}$ denotes a Coulomb propagator, which satisfies
$\Delta^{-1}\Delta=\Delta\Delta^{-1}=1$.
Then, the dual field strength can  be
rewritten as $F \!=\! d B^{\mathrm{reg}} + 2 \pi d \Delta^{-1}\!
\delta * j $.
From the square of the second term of the dual field strength,
one  gets the Coulombic interaction between electric currents, which
remains even if there is no
monopole field.
The rest of the effective action comes through
the interaction between the monopole
field and the dual gauge field.

After  the functional path-integration over $B^{\mathrm{reg}}$ and
$\eta$,  introducing 2-form Kalb-Ramond (KR) field $h$,
we obtain the action in terms of  $h$ and $\phi$.
We  divide it into three parts as 
\bea
S^{(1)} &\!=&\!
2 \pi^2 \beta_g (j,\Delta^{-1} j),
\label{eq:s-coulomb}
\\
S^{(2)} &\!= & \!
(d \phi)^2 + \frac{1}{4}(dh, 
\left \{ \frac{1}{\phi^2} - \frac{1}{v^2} \right \} d h)
+ \lambda (\phi^2 -v^2)^2,
\\
S^{(3)} &\!=& \!
\frac{1}{4 v^2}(dh)^2 \! + \frac{1}{2 \beta_g} \!
\left\{\!(h)^2\!-
(\delta h, \Delta^{-1}\! \delta h)\!\right\}
\!- 2 \pi i \!\left\{ \!(h, \Sigma)\! -\! ( \delta h,
\Delta^{-1}\! j)\!\right\}.
\eea
Here $S^{(1)}$ is the pure
Coulombic interaction from the dual field strength as discussed
above,
$S^{(2)}$ 
contains monopole modulus $\phi$,
which  plays a role 
only in the core region of the flux tube, where $\phi
\!\neq \!v$ or $d \phi \!\neq \!0$, and
$S^{(3)}$ is the rest, which remains even where $\phi \!=\! v$. 
From the structure of a classical flux-tube solution,
we know that  $\phi$ varies from $0$ to $v$ at the region 
$0 \!\leq\! \rho\! \leq\! m_{\chi}^{-1}$, where $\rho$ denotes transverse
distance from the position of the Dirac string $\Sigma$.
This region,  characterized 
by $m_{\chi}^{-1}$, is nothing but the core region.
Therefore, we consider $m_{\chi}$ as an effective cutoff and evaluate 
the action $S^{(2)} $ and $S^{(3)} $ above and below this cutoff.
The  $S^{(1)}$ does not depend on $m_{\chi}$.

To treat the ultraviolet scale with respect to the cutoff $m_{\chi}$ 
(higher resolution than $m_{\chi}^{-1}$),
one must evaluate the action $S^{(2)} +
S^{(3)}_{< m_{\chi}^{-1}}$,
 since the detailed structure of the flux tube, 
variation of $\phi$, is visible.
From the knowledge of the classical solution,
we speculate that this part gives a Nambu-Goto action with a certain string
tension $\sigma_{\mathrm{core}}$ 
and non-confining potential $S_{\mathrm{core}}(j)$.

On the other hand, for the infrared scale 
(lower resolution than $m_{\chi}^{-1}$),
one can only see the surface of the flux tube
and cannot recognize inside of it.
Thus $S^{(2)}$ gives no 
contribution 
and $S^{(3)}_{> m_{\chi}^{-1}}$ are essential
to describe the system.
In this case,  further path-integration over the KR field $h$ is possible,
which yields
\bea
S^{(3)}_{> m_{\chi}^{-1}}\!(\Sigma, j) \!&=&\!
\left[2\pi ^{2} \!\beta_{g} (j,\!\left[ D -\Delta^{-1}\! \right] \!j)
\!+\!
4\pi^{2}\!v^{2} \!(\Sigma, D\Sigma)\right]_{> m_{\chi}^{-1}},
\label{eq:low}
\eea
where $D \equiv (\Delta + m_{B}^{2})^{-1}$ is the propagator of the
massive KR field.
The last term gives another Nambu-Goto action 
with a string tension 
$\sigma_{\mathrm{surf}} = \pi v^{2}\! \ln
((m_{B}^{2}+m_{\chi}^{2})/m_{B}^{2})$
as a leading contribution of derivative expansion,
together with  corrections, for instance,
a rigidity term with the negative coefficient.
By adding all contributions,
we  arrive at  the effective string action of the DGL theory
as
\bea
&&S_{\mathrm{eff}}(\Sigma,j)
\!=\! S(j) +
 ( \sigma_{\mathrm{core}} +\sigma_{\mathrm{surf}})\!\!\int\!\!d^{2}\xi
 \sqrt{g(\xi)}+ (\mbox{corrections}),
\eea
where $\xi^{a} (a=1,2)$ parametrize the string world sheet
described by the coordinate
$\tilde{x}_{\mu}(\xi)$ and $g(\xi)$ the
determinant of the induced metric $g_{ab}\equiv 
\frac{\partial\tilde{x}_{\mu}(\xi)}{\partial \xi^{a}}
\frac{\partial\tilde{x}_{\mu}(\xi)}{\partial \xi^{b}}$.
The non-confining part of the effective action is given by
\bea
&&S(j) \!=\! 
2\pi ^{2}\! \beta_{g} (j, \Delta^{-1}\! j) + S_{\mathrm{core}}(j) +
\left[2\pi ^{2} \!\beta_{g} (j,\!\left[ D -\Delta^{-1}\! \right] \!j)
\!\right]_{> m_{\chi}^{-1}}.
\eea
We find that the Coulombic interaction from  $S^{(1)}$ 
is partially cancelled by that from  $S^{(3)}_{> m_{\chi}^{-1}}$. 
In the London limit, $m_{\chi} \!\to\! \infty$,
the complete cancellation of the Coulombic term takes place and
$S_{\mathrm{core}}(j) $ varnishes.
Thus only the  Yukawa term from $S^{(3)}$ remains.

To summarize,
we have studied the effective string action of the $U(1)$ DGL theory.
The effect of the finite thickness
appears not only in the string tension but also in
the shape of the non-confining
potential, which becomes not pure Yukawa nor pure Coulomb potential,
but something in between for  a finite monopole mass $m_{\chi}$.
This feature is the same even in the 
the $U(1) \!\times\! U(1)$ DGL theory,
which corresponds to an infrared effective theory of 
$SU(3)$ Yang-Mills theory.
In fact, an extension of these path-integral analyses
is straightforward by using the Weyl symmetric representation
of the $U(1) \!\times\! U(1)$ DGL action.\cite{Koma:2002rw}

\end{document}